\documentclass[%
aps,prl,showpacs,twocolumn,showkeys,amssymb,amsmath,superscriptaddress,reprint
]{revtex4-1}
\pdfoutput=1

\usepackage{graphicx}
\usepackage{bm}
\usepackage[utf8]{inputenc}
\usepackage[T1]{fontenc}
\usepackage{float}
\usepackage{color}
\usepackage{comment}
\usepackage{xfrac}
\usepackage[english]{minitoc}
\usepackage{multirow}
\usepackage{textcomp}
\usepackage{gensymb}
\usepackage{adjustbox}
\usepackage{caption}
\usepackage{booktabs}

\makeatletter
\let\saved@includegraphics\includegraphics
\AtBeginDocument{\let\includegraphics\saved@includegraphics}
\renewenvironment*{figure}{\@float{figure}}{\end@float}
\makeatother

\begin{document}


\title{Towards an understanding of the chemisorption and catalytic activity of Pd$_{X}$Ru$_{1-X}$ nanoparticles using photoelectron spectroscopy}
\author{Ibrahima Gueye}
\affiliation{
Synchrotron X-ray Group, Research Center for Advanced Measurement and Characterization, National Institute for Materials Science (NIMS), 1-1-1 Kouto, Sayo, Hyogo 679-5148, Japan. 
}
\affiliation{
 Synchrotron X-ray Station at SPring-8, Research Network and Facility Services Division, NIMS, 1-1-1 Kouto, Sayo, Hyogo, Japan. 
}
\author{Anli Yang}
\affiliation{
 Synchrotron X-ray Station at SPring-8, Research Network and Facility Services Division, NIMS, 1-1-1 Kouto, Sayo, Hyogo, Japan. 
}
\affiliation{
Present address : Shenzhen Institute of Wide-bandgap Semiconductors (WinS) No.1088, Xueyuan Rd., Xili, Nanshan District, Shenzhen, Guangdong, 518055, P.R.China.
}
\author{
L. S. R. Kumara
}
\affiliation{
 Synchrotron X-ray Station at SPring-8, Research Network and Facility Services Division, NIMS, 1-1-1 Kouto, Sayo, Hyogo, Japan. 
}
\affiliation{
Japan Synchrotron Radiation Research Institute (JASRI), 1-1-1 Kouto, Sayo, Hyogo 679-5148, Japan.
}

\author{
Satoshi Hiroi
}
\affiliation{
Synchrotron X-ray Group, Research Center for Advanced Measurement and Characterization, National Institute for Materials Science (NIMS), 1-1-1 Kouto, Sayo, Hyogo 679-5148, Japan. 
}
\affiliation{
 Synchrotron X-ray Station at SPring-8, Research Network and Facility Services Division, NIMS, 1-1-1 Kouto, Sayo, Hyogo, Japan. 
}

\author{Okkyun Seo
}
\affiliation{
Synchrotron X-ray Group, Research Center for Advanced Measurement and Characterization, National Institute for Materials Science (NIMS), 1-1-1 Kouto, Sayo, Hyogo 679-5148, Japan. 
}
\affiliation{
 Synchrotron X-ray Station at SPring-8, Research Network and Facility Services Division, NIMS, 1-1-1 Kouto, Sayo, Hyogo, Japan. 
}

\author{Jaemyung Kim
}
\affiliation{
Synchrotron X-ray Group, Research Center for Advanced Measurement and Characterization, National Institute for Materials Science (NIMS), 1-1-1 Kouto, Sayo, Hyogo 679-5148, Japan. 
}
\affiliation{
 Synchrotron X-ray Station at SPring-8, Research Network and Facility Services Division, NIMS, 1-1-1 Kouto, Sayo, Hyogo, Japan. 
}

\author{Kohei Kusada
}
\affiliation{
Division of Chemistry, Graduate School of Science, Kyoto University, Kitashirakawa Oiwake-cho, Sakyo-ku, Kyoto 606-8502, Japan.  
}

\author{Hiroshi Kitagawa
}
\affiliation{
Division of Chemistry, Graduate School of Science, Kyoto University, Kitashirakawa Oiwake-cho, Sakyo-ku, Kyoto 606-8502, Japan.  
}

\author{Osami Sakata
}
\email{SAKATA.Osami@nims.go.jp}
\thanks{Corresponding author}
\affiliation{
Synchrotron X-ray Group, Research Center for Advanced Measurement and Characterization, National Institute for Materials Science (NIMS), 1-1-1 Kouto, Sayo, Hyogo 679-5148, Japan. 
}
\affiliation{
 Synchrotron X-ray Station at SPring-8, Research Network and Facility Services Division, NIMS, 1-1-1 Kouto, Sayo, Hyogo, Japan. 
}
\affiliation{
Department of Materials Science and Engineering, Tokyo Institute of Technology, Nagatsuta, Yokohama, Japan. 
}


\begin{abstract}
Chemisorption process and catalytic activity of CO to CO$_{2}$ conversion on Pd$_{X}$Ru$_{1-X}$ (X : 0, 0.1, 0.3, 0.5, 0.7, 0.9 and 1) extended as a benchmark for further investigation of very complex nanoparticle (NP) structures have been checked into thoroughly by using the synchrotron-based hard X-ray photoelectron spectroscopy (HAXPES). Assessment and diagnostic of core levels and valence bands data highlight valuable information regarding the surface interaction and structure of PdRu bimetallic nanoparticles (BM-NPs). Core level shift observed from the C 1s clearly emphasized that the Pd$_{0.5}$Ru$_{0.5}$ NPs which provide the highest catalytic efficiency (CO oxidation) exhibits a preferential adsorption of the CO molecule at the top site. Otherwise, our results also display that remaining NPs exhibit two components assigned to the bridge and hollow CO sites. Combination of Pd 3d$_{5/2}$ and Ru 3p$_{3/2}$ core levels results points out a transition from alloy structures to core-shell configuration with increasing the molar ratio of Ru. Valence-band maximum (VBM) demonstrates a position dependent to the molar composition of Ru. Additionally, by digging into the electronic structures (shape and width) of each NPs very deeply, we clearly emphasise the valuable relationship between composition, atomic arrangement, d-band and catalytic properties of bimeatllic PdRu. According to the HAXPES results, the strong and clear evidence for the enhanced CO to CO$_{2}$ conversion ability of Pd$_{0.5}$Ru$_{0.5}$-NP may be assigned to the existence of a most favourable spatial d-band extension and an optimum balance between the core-shell and alloy structures. Electronic and chemical data and interpretations are consistent with the PdRu catalytic performances. This present prospection gives a new step towards a better knowledge on the catalytic surface interaction of binary, ternary and multimetallic NPs with reactants.
\end{abstract}

\maketitle



Over 80 $\%$ of manufactured products from the chemical industry include catalytic processes in the course of their fabrication. Catalysis consists in the acceleration of reactions through the use of substances called catalysts. Catalysts is not consumed during the reaction and highly decreases the energetic cost of reactions. The performances of a catalyst are generally described in terms of activity, selectivity and stability \cite{Mostafa2009, Abad2008, Li2001}. Catalysis by transition metal nanoparticles (NPs) can generally provide several ways to improve the rate at which the reactants are consumed or the products are formed. Otherwise, even though single metal shows a good performances, however the maximum catalytic activities from these transition metals is far from being achieved \cite{Gasteiger2009}. 

Based on the possibilities offered by these materials, ongoing extensive researches on the combination of several single metal based on the Pt-group were started in order to optimize the efficiency of NPs. Compared to monometallic nanoparticles (M-NPs), investigation of bimetallic nanoparticles (BM-NPs) have attracted huge attention not only for the automotive exhaust converter, methane combustion system, hydrogen storage and CO-oxidizing abilities \cite{Weskamp1998, Yamaguchi2003, Jacobsen2000, Perkas2011, Bowker2008}, but also for the fundamental understanding of the catalytic surface synergies \cite{Singh2013, Kim2014, Jiang2011, Mott2007}. In fact, mixture of two metals may display not only the combination of the properties related to the presence of two mono-metals, but also new properties due to a synergy between two metals \cite{Toshima1998}. 

From these myriad possible combinations, Kitagawa and co-workers highlighted a similar hydrogen storage property between Ag$_{0.5}$Rh$_{0.5}$ and Pd while the two individual metals (Ag and Rh) have no hydrogen storage ability \cite{Kusada2010}. These enhanced properties in BM-NPs are confirmed by results from the mixture Ru and Pd metals \cite{Kusada2014}. Wu \textit{et al.} \cite{Wu2014} and Kusada \textit{et al.} \cite{Kusada2014} reported respectively that the composition activity studies on Pd$_{0.6}$Ru$_{0.4}$ and Pd$_{0.5}$Ru$_{0.5}$ NPs exhibit enhanced catalytic activities compared with Ru and Pd M-NPs under the same condition. 

Furthermore, even though superior catalytic performances have been demonstrated in a large part of BM-NPs like Pd-Ru, Pd-Au, Ru-Cu, Pt-Ru, Ag-Ru and Au-Pt \cite{Wu2016, Sasaki2012, Liu2004, Alayoglu2008, Adekoya2015, Esfahani2010}, there is a huge lack of fundamental researches about the clear understanding of the BM-NPs surface chemisorption mechanisms. Catalytic surface efficiencies obtained from the combination of two different metals have been related distinctively either to the geometric structure (strongly related to the Madelung constant of crystal) or electronic structure (considerably dependant to the average valence electron binding energy of different elements in interactions). As already reported by Song \textit{et al.} \cite{Song2019}, geometric parameters can be tuned by adjusting composition, atomic ordering, morphology and size while the desired electronic properties of a target monometal can be reproduced via the mixing of other elements. However, a careful inspection of the BM-NP properties suggests that catalytic activities of BM-NPs do not only result neither to the geometric nor to the electronic structures but also a combination of both parameters as suggested by Kim et \textit{al.} \cite{Kim2014} and Guidez et \textit{al.} \cite{Guidez2012}. Due to the high complexity involved in evaluating the factors governing the chemisorption mechanisms, we will attempt to anticipate the growth needs of the fundamental understanding of mixture NPs. 

\begin{figure}[!ht] 
\begin{flushleft} 
\includegraphics[width=0.48\textwidth]{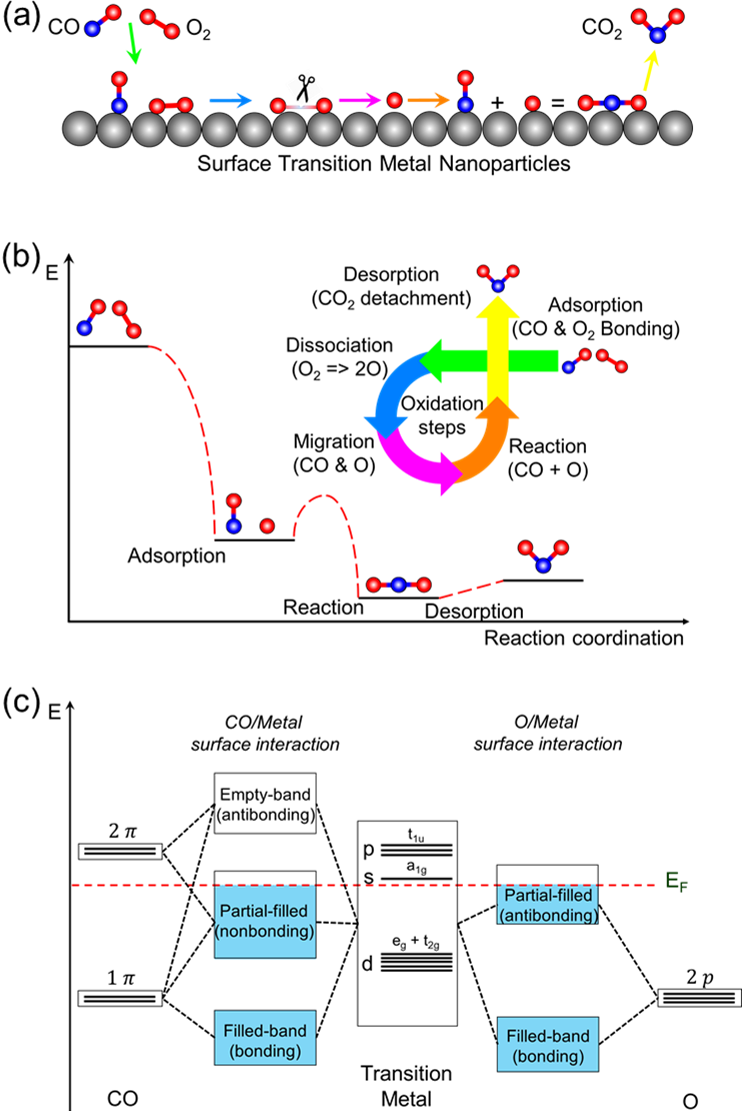} 
\captionsetup{justification=raggedright}
\captionof{figure}
{
(a) Schematic of the Langmuir-Hinshelwood mechanism drawing the adsorption of CO and O$ _{2} $ species, the diffusion over the metal surface and the reaction and production of CO$ _{2} $, (b) illustrative potential energy diagram from CO and O$ _{2} $ species to CO$ _{2} $ product and (c)  Molecular orbitals diagram for CO-Metal and O-Metal bonds.
} 
\label{figure1}
\end{flushleft} 
\end{figure}
Prototypical bimetal based on Pd and Ru metals have been selected to study the intriguing relation between the chemisorption activities and the intrinsic properties. The choice PdRu NP is motivated by the fact that Pd and Ru metals can be useful from a strategic point of view. In fact, Ru-NP has attracted much attention as an effective catalyst for the steam-reforming reaction of methane, which is the main component of shale gas \cite{McFarland2012, Lee2012}. Pd-NP shows an outstanding response for hydrogen storage, fuel-cell electrodes and exhaust-gas purification despite the catalytic activity of Pd is constrained by surface poisoning by CO \cite{Xiao2009, Newton2007, Schultz2006}. 

Because of the numerous role of Pd and Ru NPs in many industrial catalytic processes, it seems to be essential to find out what is the bunch of critical parameters governing the electrochemical performances. In this context, we will tend to correlate the catalytic performance to the chemical and electronic structures of the Pd$_{X}$Ru$_{1-X}$-NPs with different compositions  by using photoelectron spectroscopy based on the synchrotron source (HAXPES). In fact, electronic and chemical properties of these transition metal combinations arise mainly from the bonding behavior of metal d-band electrons interacting with reactants. Catalytic CO oxidation on transition metal surfaces described by the Langmuir-Hinshelwood mechanism can be a good way to theoretically highlight the degree of complexity of the chemisorption process. It can also serves as a link to understanding the electronic and chemical behavior during the adsorption of reactants such as CO and O$_{2} $ diatomic molecules. 

From \textcolor{black}{Fig.\ref{figure1}(a)}, we can follow a sequence of three steps: (i) adsorption of CO  and O$ _{2} $, (ii) scission of O$ _{2} $ and diffusion of the species over the surface until the reactants CO and O meet and generate the CO$ _{2} $ product and (iii) desorption of the CO$ _{2} $ product \cite{Gueye2019}. Concerning the energetic point of view \cite{Chorkendorff2003}, the catalytic interaction begins with a bonding of CO and O$ _{2} $ on the transition metal surface as reported schematically in \textcolor{black}{Fig. \ref{figure1}(b)}. This adsorption step is exothermic \cite{Allian2011}, and the free energy is lowered. There then follows the reaction between CO and O$ _{2} $ while they are bound to the catalyst. This step is associated with an activation energy. Finally, the CO$ _{2} $ product releases from the transition metal surface in an endothermic step. Note that from \textcolor{black}{Fig. \ref{figure1}(b)}, the highest barrier controls the kinetics of the overall process.

Furthermore, though metal substrate does not visually participate as a reactant nor as a product and left unchanged when the CO$ _{2} $ is came off, however the reaction could not happen without the electronic exchange occurred between reactants and the surface metal. In fact, when the valence band interaction between diatom molecules (CO or O$ _{2} $) and metal is too weak, It will be impossible to create the CO$ _{2} $ due to the low stability of reactant on the metal surface. In the same way when one or both reactants are strongly bonded (poising effect) with the surface catalyst, any CO$ _{2} $ product could be obtain. Intermediate electronic exchange is always needed in order to achieve a good catalysis process. Further, as we can see in \textcolor{black}{Fig. \ref{figure1}(c)}, atomic and molecular orbital results when CO and O$ _{2} $ are brought toward a transition metal surface are totally different due to the diverse types of interacting orbitals (azimuthal quantum number) and their symmetries (magnetic quantum number) \cite{LaRue2015}. In case of CO,  interaction mainly result to the creation of lowest, middle and highest orbitals, which correspond to the filled bonding orbital, partial filled non-bonding orbital and empty antibonding orbitals, respectively. Conversely, the O-metal hybridization leads only to the creation of bonding and anti bonding orbitals.  According to this strong implication of electronic states on the surface chemisorption, it would be helpful to have an understanding of the local electronic configuration when we design devices based on transition metal NPs for an specific catalytic applications.

In the present paper, samples synthesized by chemical reduction and investigated by HAXPES will be discussed. We highlight the preferential adsorption site of the CO molecule relative to the Pd$_{X}$Ru$_{1-X}$-NPs composition. Through the Pd 3d$_{5/2}$ and Ru 3p$_{3/2}$ core levels, chemical interactions and structural configurations will be discussed and different models of configuration will be proposed. Analytical description is also provided in order to understand the intrinsic electronic exchange into the bimetallic structure (Pd and Ru) and how this electronic charge is distributed after contact. This latter part will give us an idea about how the Pd-Ru electronic exchange influences the properties of these systems. Finally, electronic structure investigation via the valence band (VB) will be examined and the electron distribution determined. 
\section*{Results and discussion}
\begin{figure}[!ht]
\centering
\includegraphics[width=0.48\textwidth]{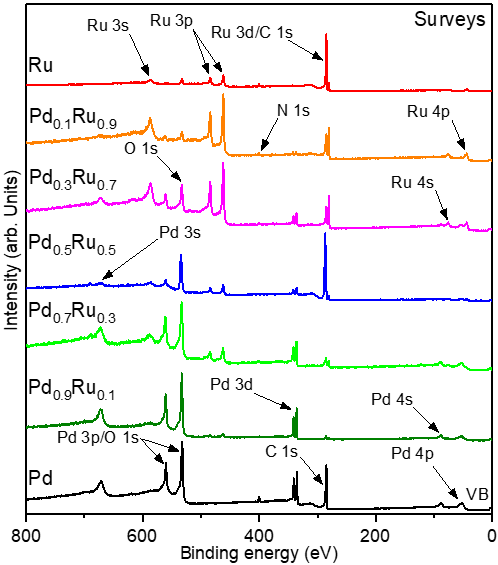} 
\captionsetup{justification=raggedright}
\captionof{figure}{Survey spectra of Pd-NP (black), Pd$_{0.9}$Ru$_{0.1}$ (green), Pd$_{0.7}$Ru$_{0.3}$ (olive), Pd$_{0.5}$Ru$_{0.5}$ (blue), 
Pd$_{0.3}$Ru$_{0.7}$ (magenta), Pd$_{0.1}$Ru$_{0.9}$ (orange)  bimetallic NPs and Ru-NP (red). All expected elements (Palladium, Ruthenium, Carbon, Oxygen and Nitrogen) are observed.} 
\label{figure2}
\end{figure}

\indent \\
The HAXPES measurements over an extended energy (from 0 to 800 eV) were carried out in order to assess each NPs in terms of elemental composition and relative abundances. These typical survey spectra reported in \textcolor{black}{Fig. \ref{figure2}} show the presence of : nitrogen (N 1s) from the PVP surround; the key elements Pd and Ru from mono and bi metallic NPs; C and O from the carbon nanopowder, PVP protecting agent, water and adsorbed molecules (from the ambient air). From the Pd-NP to the Pd$_{0.9}$Ru$_{0.1}$ spectra, the relative concentration of the Pd 3d$_{5/2}$ signals undergo an continual attenuation which is consistent with the nominal composition trend. Same progression is followed by the Ru 3p$_{3/2}$ peak when we go from the Ru-NP to the Pd$_{0.1}$Ru$_{0.9}$ spectra. Higher-resolution spectra of Pd 3d$_{5/2}$ and Ru 3p$_{3/2}$ regions as well as the valence band section of respective samples were taken and will be discussed in the following parts of this manuscript.
\begin{figure}[!ht]
\centering
\includegraphics[width=0.48\textwidth]{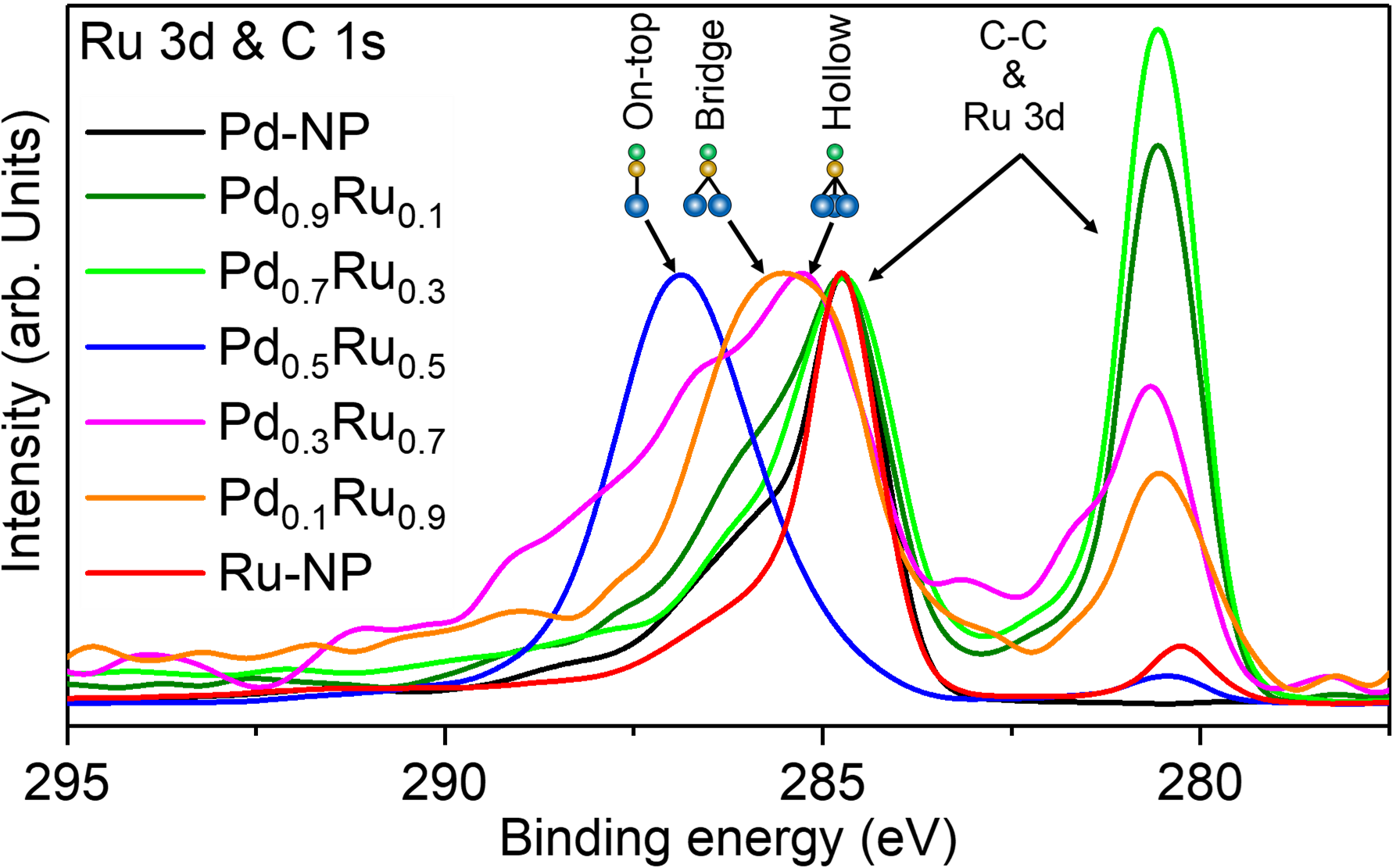} 
\captionsetup{justification=raggedright}
\captionof{figure}
{Zooming on the Ru 3d and C 1s range (from the survey spectra). In order to compare the CO adsorption, we normalize all spectra relatively to the maximum intensity to the C 1s levels instead of the Ru 3d$_{5/2}$ levels. The binding energy position of the maximum intensity of the C 1s component show a composition-dependence. Pd$_{0.5}$Ru$_{0.5}$ displays a hugh binding energy shift (C 1s) relative to others. Blue, yellow and green balls are attributed to metal, carbon and O respectively.} 
\label{figure3}
\end{figure}

\textcolor{black}{Figure \ref{figure3}} reports C 1s and Ru 3d binding energy values as the function of the composition of PdRu-NPs. Note that, normalization of spectra based on the C 1s maximum intensity have been made in order to clearly highlight the CO preferential surface localization (bond). Binding energy values from C 1s allow us not only to distinguish CO or C-C (C sp$^{2}$, C sp$^{3}$) component but also differentiate the different location of CO molecular on NP surfaces. In fact, besides peaks around 280 eV and 284.5 eV which are found corresponding to Ru 3d$_{5/2}$ and C-C/Ru 3d$_{3/2}$, respectively, we can underline several possibilities of CO to interact with PdRu-NPs. Binding energies from numerous studies have been used to identify specific CO sites such as terminal (on-top), bridge and hollow \cite{Kaichev2003, Naitabdi2018}. The peak around 285.2 eV is recognized as corresponding to CO adsorbed in hollow sites on the metallic surface. Peak around 285.6 eV is assigned to CO binding in the bridge site. Finally, around 286.9 eV, CO component is attributed to CO adsorbed on the on-top site.

By combining data from these latter observations (\textcolor{black}{Fig. \ref{figure3}}) and from Kitagawa team reports on these PdRu-NPs \cite{Kusada2014}, we can highlight that conversion of the CO to CO$_{2}$ is strongly affected by the adsorption sites of CO molecules. The importance of bonding (adsorption) sites on the catalytic activity of the co-adsorption system of mCO plus nO$_{2}$ on Ru(0001) were clearly and theoretically described by Schiffer \textit{et al.} \cite{Schiffer1997}, Narloch \textit{et al.} \cite{Narloch1995} and Kostov \textit{et al.} \cite{Kostov1992}. They also point out contributions from mutual interactions of reactants and provide different energetically more favourable co-adsorption configuration.

Below 50 \% of Ru in the Pd$_{X}$Ru$_{1-X}$ composition in \textcolor{black}{Fig. \ref{figure3}}, main contribution of the C 1s suggests a preferential adsorption of C (C sp$^{2}$ or C sp$^{3}$) on the NP samples. Additionally, the presence of a partial mixture of hollow- and bridge-bonded CO cannot be fully excluded due to the appearance of the low shoulder near the C-C peak. Furthermore, Pd$_{0.5}$Ru$_{0.5}$ NP shows a different CO bond type with mainly on-top adsorption. Finally, for Pd$_{0.3}$Ru$_{0.7}$ and Pd$_{0.1}$Ru$_{0.9}$, the CO conversion is influenced by the combination of a hollow- and bridge-bonded. Following \textcolor{black}{Fig \ref{figure3}}, we highlight a competition between  the C (C sp$^{2}$ or C sp$^{3}$) adsorption and the CO molecule as function of the molecular ratio of Pd and Ru. However, these results show only the presence of the CO on the PdRu surface but does not indicate on which atoms (Pd or Ru) the CO is bound.

\begin{figure*}[!ht]
\centering
\includegraphics[width=0.98\textwidth]{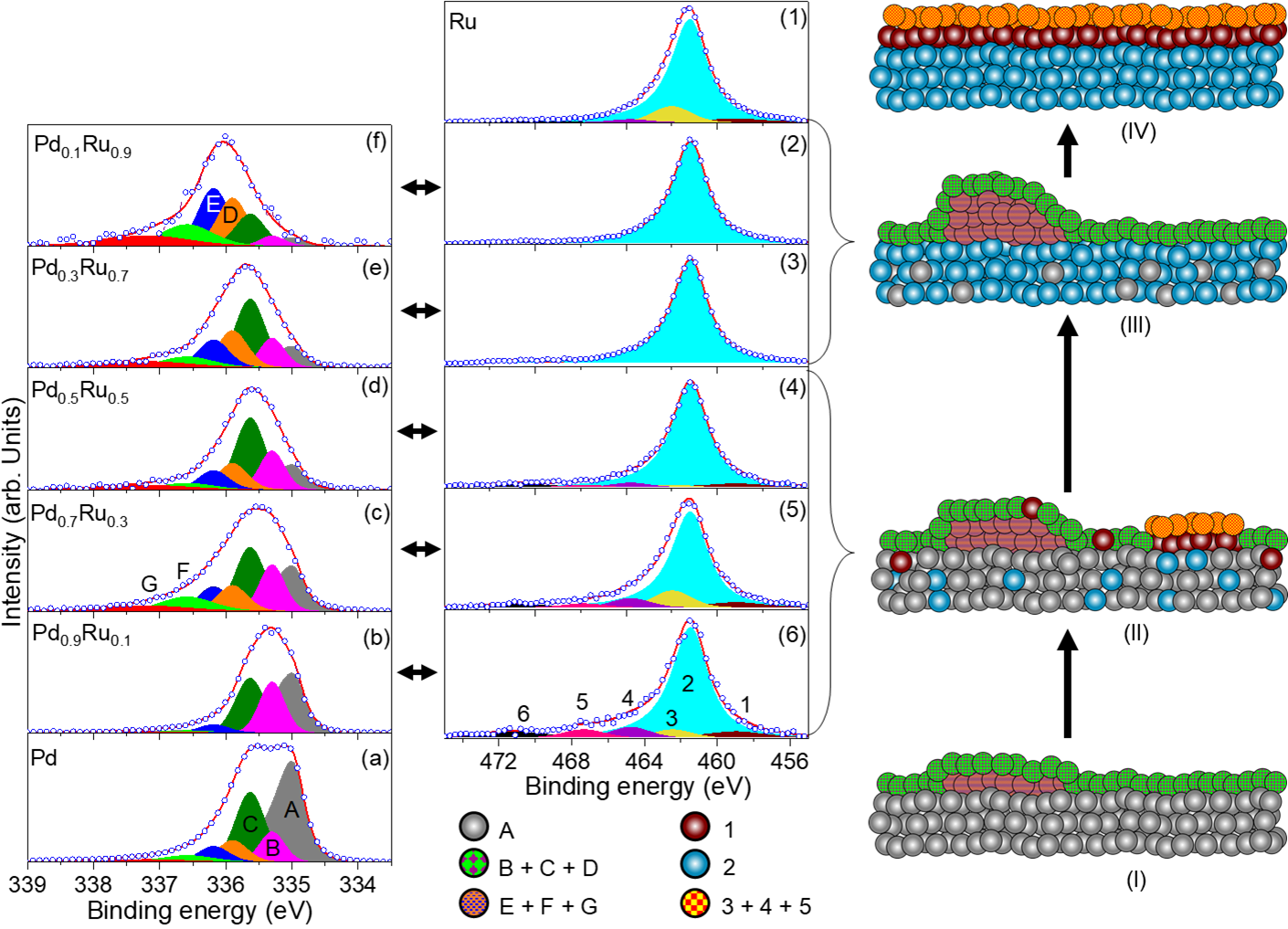} 
\captionsetup{justification=raggedright}
\captionof{figure}
{HAXPES data summary of Pd 3d$_{5/2}$ (from (a) to (f) see left-hand panel) and Ru 3p$_{3/2}$ (from (6) to (1) shown in the center panel) spectra with fits. From the bottom up : (a)Pd-NP, (b) and (6) Pd$_{0.9}$Ru$_{0.1}$, (c) and (5) Pd$_{0.7}$Ru$_{0.3}$, (d) and (4) Pd$_{0.5}$Ru$_{0.5}$, (e) and (3) Pd$_{0.3}$Ru$_{0.7}$, (f) and (2) Pd$_{0.1}$Ru$_{0.9}$ and (1) Ru-NP, respectively. For Pd 3d$_{5/2}$ spectra : A, B, C, D, E, F and G components have been assigned to Pd$^{0}$, Pd-C, Hollow CO \& Bridge CO, On-top CO, 4-fold Pd, PdO, PdO$_{2}$, respectively. For Ru 3p$_{3/2}$ core levels, (1) is attributed to the surface unsaturated contribution, component 2 is linked to the metallic Ru (Ru$^{0}$). Components 3, 4 and 5 are related to the Ru shell oxidation Ru$^{4+}$, Ru$^{6+}$ and Ru$^{8+}$, respectively. Peak 6 is from a satellite component Ru$_{satellite}$. The models I, II, III and IV reported in the right-hand panel illustrate the origin of the different contribution in the Pd 3d$_{5/2}$ and Ru 3p$_{3/2}$ core levels.} 
\label{figure4}
\end{figure*}
\textcolor{black}{Figure \ref{figure4}} displays the Pd 3d$_{5/2}$ and Ru 3p$_{3/2}$ HAXPES results following Ru molar ratio. Starting with the fcc Pd-NPs data in \textcolor{black}{Fig. \ref{figure4}(a)}, we remark that even though the Pd$^{0}$ component (peak A) intensity is very important, Pd 3d$_{5/2}$ also exhibits components with a core level shifted at higher binding energy. The presence of these latter reflects that either an oxidation and/or a C (C sp$^{2}$ or C sp$^{3}$) phase around the Pd atoms. Components B, C are related to the Pd-C-C (called Pd-C in the following text) and the Pd-CO located in the hollow/bridge \cite{Kaichev2003, Westerstrom2011, Toyoshima2012}, respectively. This result is consistent with the C 1s component of Pd-NP (black line) in \textcolor{black}{Fig \ref{figure3}} because CO (hollow/bridge) contribution  is higher than the C (C sp$^{2}$ or C sp$^{3}$ contribution. This situation can be explained by the fact that the C-C peak overlap with the Ru 3d$_{3/2}$, which increases the intensity of the peak at 284.5 eV. 

\textcolor{black}{Figures \ref{figure4}(b-f)} show a decrease of the Pd$^{0}$ (peak A) and the Pd-C component (peak B) while D (On-top CO) and E (4-fold Pd) contribution increases \cite{Kaichev2003, Westerstrom2011, Toyoshima2012}. Low amounts of F (PdO) and G (PdO$_{2}$) peaks \cite{Kaichev2003, Westerstrom2011, Toyoshima2012} can be also from the oxidation progress into the bulk of NPs. The relative contribution of the component C stays almost stable except in the Pd$_{0.1}$Ru$_{0.9}$.

Conversely to the Pd 3d$_{5/2}$, Ru 3p$_{3/2}$ in \textcolor{black}{Fig. \ref{figure4} (1-6)} indicate that the major part of the Ru remains metallic (Ru$^{0}$) according to component 2. We also observe that below 50 $ \% $ of Ru molar ratio, we have three additional components at high binding energy \cite{Morgan2015, Hota2019, Kruefu2016, Lizzit2009} corresponding to the Ru$^{4+}$ (peak 3), Ru$^{6+}$ (peak 4) and Ru$^{8+}$ (peak 5), respectively. Clean surface contribution (unsaturated as dangling bonds) corresponding to peak 1 is also observed in these latter samples. For PdRu-NPs with 70 $ \% $ and 90 $ \% $ of Ru, we record only the Ru$^{0}$ (peak 2).

Results from \textcolor{black}{Fig. \ref{figure4}} highlight that metal segregation or miscibility in PdRu-NPs depends mainly on the molar composition. PdRu NPs extended surfaces have been shown to exist in three distinct surface structures. The surface configurations of these phases are schematically illustrated in \textcolor{black}{Fig. \ref{figure4} (I-IV)} and can be named as follows \cite{Andersson2009, Knudsen2007, Nierhoff2015}: the Pd overlayer on Ru in model III, the bi-metal alloy in model II and the mono-metal surface adsorption in models I and IV.
\begin{table}[!ht]
\centering
\caption{Experimentally determined core level binding energy and components for Pd 3d$_{5/2}$ and Ru 3p$_{3/2}$ from Pd$_{X}$Ru$_{1-X}$ (X : 0, 0.1, 0.3, 0.5, 0.7, 0.9 and 1) NPs \cite{Kaichev2003, Westerstrom2011, Toyoshima2012, Morgan2015, Hota2019, Kruefu2016, Lizzit2009}.}
\label{At_3}
\begin{adjustbox}{width=0.48\textwidth}
\begin{tabular}{cc c cc}
\toprule
\midrule 
\multicolumn{2}{c}{Pd 3d$_{5/2}$} & & \multicolumn{2}{c}{Ru 3p$_{3/2}$} \\
\cmidrule(l){1-2}\cmidrule(l){3-5}
 Peaks (eV)  &components  & &  Peaks (eV)&  components   \\
\midrule
A (335.0)&   Pd$^{0}$   & &  1 (458.9) & Surface  \\
B (335.3)&  Pd-C \& 2-fold Pd  & &  2 (461.5) & Ru$^{0}$ (Bulk)  \\
C (335.6)&  Hollow CO \& Bridge CO  & &  3 (462.5) & Ru$^{4+}$ \\
D (335.9)&  On-top CO  & &  4 (464.5) & Ru$^{6+}$  \\
E (336.2)& 4-fold Pd   & &  5 (467.3) & Ru$^{8+}$  \\
F (336.6)&  PdO   & &  6 (471.3) & Ru$_{satellite}$ \\
G (337.2)&  PdO$_{2}$   & &  - & -  \\
\midrule 
\bottomrule
\end{tabular}
\end{adjustbox}
\end{table}

We remark that from Pd$_{0.9}$Ru$_{0.1}$ to Pd$_{0.5}$Ru$_{0.5}$, both metals are partially oxidized as we can see in the schema in \textcolor{black}{Fig. \ref{figure4} (II)}. This observation can be explained by the fact that the final structure of these latter (from the Pd$_{0.9}$Ru$_{0.1}$ to Pd$_{0.5}$Ru$_{0.5}$-NPs) is an alloy structure \cite{Alayoglu2009, Miyakawa2014}. Above Pd$_{0.5}$Ru$_{0.5}$, we observe a change from alloy to core-shell structure \cite{Alayoglu2009, Miyakawa2014} as seen in \textcolor{black}{Fig. \ref{figure4} (III)}. This assumption is consistent by the fact that the Ru is fully metallic while the Pd is almost entirely oxidized. Surface adsorption are observed on the Pd and Ru M-NPs.

Core-level analyses from \textcolor{black}{Fig. \ref{figure4}} support the fact that the best capability of the catalytic activities in the PdRu-NPs requires an optimum balance between the core-shell and alloy structures. In fact, from Kusada et \textit{al.} report \cite{Kusada2014}, the CO to CO$_{2}$ conversion temperature in all of the PdRu-NPs are lower than those in pure Ru or Pd-NPs. The CO to CO$_{2}$ conversion temperature reached a minimum (temperature) with the Pd$_{0.5}$Ru$_{0.5}$ which is located on the border between the PdRu with core-shell structures and PdRu with alloy configurations. 

Optimal performance of Pd$_{0.5}$Ru$_{0.5}$ may also be partially ascribed to the Ru atom in the interface layer situating between the bulk metallic NPs and the surface oxide layer (Ru atom with unsaturated bond). These near surface Ru metal can significantly drive or dictate the preferential occupation sites of the CO and O$_{2}$ adsorbate molecules \cite{Andersson2009, Knudsen2007, Nierhoff2015} due to their excess charge relative to the dangling bond. Furthermore, to better explain the catalytic activities of these PdRu-NP systems, it is not enough to make a good interpretation considering the effect from the structure and chemical adsorption; it is also necessary to make a more detailed analysis taking into account specific electronic interactions related to the composition.

\begin{figure}[!ht]
\centering
\includegraphics[width=0.45\textwidth]{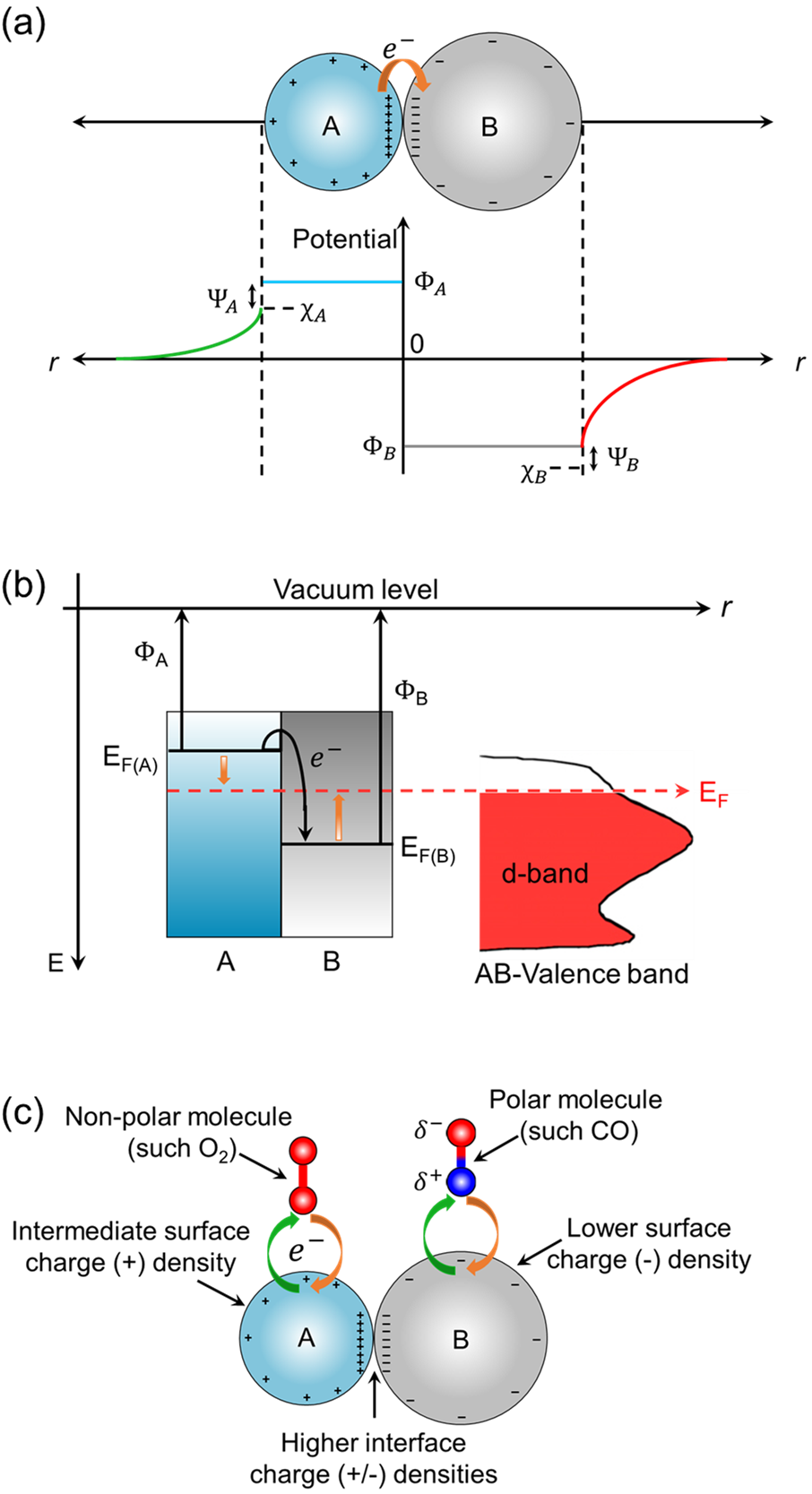} 
\captionsetup{justification=raggedright}
\captionof{figure}{Geometrical illustrations depicting the electron exchange process theory in BM-NPs: (a) shows schematically the potential distribution after metallic contact of A and B atoms. $ r $ represents the symmetry axis through the metal contact. $\Phi$, $ \Psi $ and $ \chi $ are the work function, contact potential and surface potential, respectively. (b) highlights electron flow from the lower $\Phi_{A}$ to the higher $\Phi_{B}$ upon the contact, the Fermi level ($ E_{F} $) of the equilibrate $AB$ NP and the final d-band of AB bimetal after electron filling and orbital overlap of monometal d-bands and (c) show a selective reactants (molecules) metal NP due to the none uniform distribution of the surface charge (surface potential). $\delta^{+}$ and $\delta^{-}$ represent the slightly positive and negative surface charge, respectively} 
\label{figure5}
\end{figure}
\textcolor{black}{Figure \ref{figure5}} provides an analytical description about electron exchange processes that occur from one metal to the other and the charge transferred between a metal and a reactant. We also provide details on how the exchange of electrons can be related to the chemisorption process and catalytic activity of NPs. In fact, Fermi level position (and the distribution of electrostatic potential) is tightly associated to the interfacial contact of metals. From the diagram in \textcolor{black}{Fig. \ref{figure5} (a)}, we highlight the electron flow process across the interface of two different metallic atoms (schematize by a circle) due to the contact potential difference generated upon contact. In fact, by considering two neutral A and B atoms with different radii, the potential difference between the two metals is explicitly got through the work function difference ($\Phi_{A}-\Phi_{B}$) \cite{Peljo2016, Kumsa2016}. Further, the total amount of the transferred charge will be related to the size of the contact interface. This electronic exchange (from the metal with the lower work function to the other material) allows to keep the Fermi levels ($ E_{F} $) of both metallic atoms equal as illustrated in \textcolor{black}{Fig. \ref{figure5} (b)}. In this context, the energy to extract one electron has to be the same in any surface point of the bimetallic structure. The contact potential ($ \Psi $) of each atom is constant due to the fact that most of this transferred charge is retained at the interface as an interfacial dipole \cite{Peljo2016, Kumsa2016}. Furthermore, the surface charge distribution is not uniform on each atom surface after the Fermi level equilibration. Due to the inhomogeneous surface potential ($ \chi $), metal B acquires a low surface charge density because of its large radius whereas metal A has a high surface charge density (relative to its small radius) even though the absolute charges on both metal are the same as shown in \textcolor{black}{Fig. \ref{figure5} (a)}. Consequently, there are different electrostatic interaction (attractive or repulsive) between each metal and reactants (\textcolor{black}{Fig. \ref{figure5} (c)}), which strongly affect the chemisorption mechanisms.

These theoretical predictions may contributed to explain the experimental data (in \textcolor{black}{Fig. \ref{figure4}}). Indeed,  Pd may be identified to the B and the Ru to the A in \textcolor{black}{Fig. \ref{figure5} (c)}. From that, we can show that Pd (with $ r_{Pd} $ = 1.37 {\AA} and $ \Phi_{Pd}$ $\sim$ 5.12 eV) is more ability to react with CO molecules than the Ru (with $ r_{Ru} $= 1.34 {\AA} and $ \Phi_{Ru}$ $\sim$ 4.71 eV). In fact, From the Langmuir-Hinshelwood (LH) theory of adsorption, the description of carbon monoxide oxidation (CO + \sfrac{1}{2} O$_{2}$ $\rightarrow$ CO$_{2}$) shows that the CO molecule bonds on the NP surface through the carbon atom which is ($ \delta^{+} $) due to the an unequal sharing of electrons between the carbon and the oxygen atoms (electronegativity difference between oxygen and carbon ). In this context, the negative electron density on the surface of the Pd will be the most favourable site for the carbon NP interaction.  We assume also that the O$_{2}$ will tend to link with the Ru atoms according to the higher oxidation observed in the \textcolor{black}{Fig. \ref{figure4}(b)}. Indeed, for O$_{2}$ the electron density around each oxygen atom is equal which indicates that the bond's pair of electrons is shared equally between the two oxygen atoms. However, non-bonding pairs of electrons (or lone pair) may tend to interact with the surface of Ru which is relatively electro-positive. This distinguish adsorption site for CO and O$_{2}$ represents a good way to reduce the overcrowding issue on the surface. It also facilitate the migration of CO and O (after dissociation) and leads to the efficiency formation of the CO$_{2}$.

\begin{figure}[!ht]
\begin{flushleft} 
\includegraphics[width=0.49\textwidth]{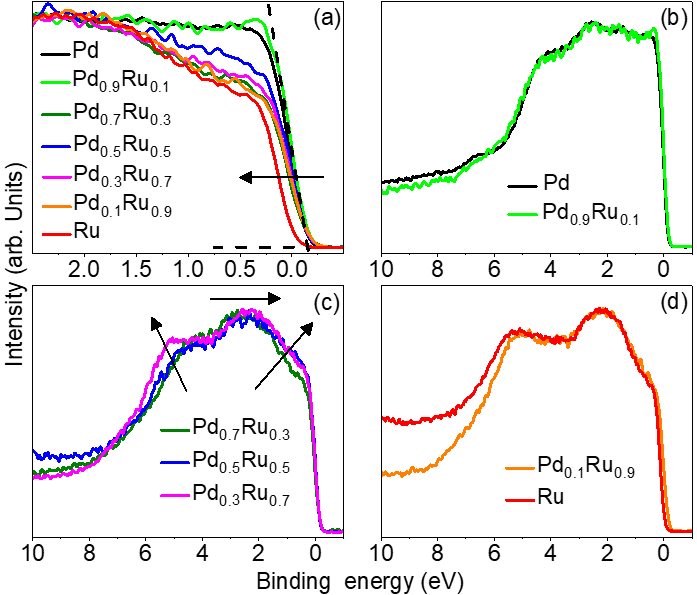}
\captionsetup{justification=raggedright} 
\captionof{figure}{Valence band spectra from HAXPES from Pd$_{X}$Ru$_{1-X}$-NPs are treated and reported: (a) shows the valence band maximum depending on the composition of the bimetals. (b) Pd-NP (Black), Pd$_{0.9}$Ru$_{0.1}$, (c) Pd$_{0.7}$Ru$_{0.3}$ (olive), Pd$_{0.5}$Ru$_{0.5}$ (blue), Pd$_{0.3}$Ru$_{0.7}$ (magenta) and (d) Pd$_{0.1}$Ru$_{0.9}$ (orange), Ru (red) highlight the shapes and widths of all samples, respectively.} 
\label{figure6}
\end{flushleft} 
\end{figure}

\textcolor{black}{Figure \ref{figure6}} shows the experimental comparison of electronic states of Pd$_{X}$Ru$_{1-X}$ NPs particularly on the width, shapes and position of the valence band maximum (VBM). Linear extrapolation of the valence \cite{Gueye2016} reported in \textcolor{black}{Fig. \ref{figure6} (a)} gives valuable information. We observe that the VBM of Ru-NP move toward the higher binding energy relative to the Pd-NP. These results are consistent with the fact that the Pd-NP presents mainly Pd-CO and Pd-C adsorption whereas the Ru-NP displays several high degree of oxidation (Ru$^{4+}$, Ru$^{6+}$ and Ru$^{8+}$) as reported in \textcolor{black}{Fig. \ref{figure4}}. In other word the electronic charges exchange (or interaction) involved in the Pd-CO and Pd-C are less than in the case of Ruthenium in which  the Ru surface atoms bound with several oxygen atoms. Finally, VBM positions of  Pd$_{X}$Ru$_{1-X}$-NPs are in agreement with of the combination of the Ru and Pd monometallic properties. Furthermore, besides the charge transfer already illustrated in \textcolor{black}{Fig. \ref{figure5}} and \textcolor{black}{Fig. \ref{figure4}}, some other parameters can have a drastic effect on catalytic responses of NPs.

Catalytic activity can be explained partially in terms of lattice constant size. In fact, when the lattice constant decreases (increases), the spatial delocalization of the orbitals increases (decreases) producing the extension (compression) of d band. In additional we assume that sizes of orbitals involved to these interactions may probably undergo some modifications. Thus, the adsorption strength of reactants on metal surface also changes due to the modifications of the electronic valence states. In fact, based on the analytical Newns-Anderson-Grimley model \cite{Newns1969Grimley1971} to describe the chemisorption, we can remark that the spatial delocalization of orbitals (related to the lattice constant) impacts the interaction between reactants and d band structures of metal surface. In the case of narrow valence band like d band, bonding and antibonding rehybridized states usually result due to the fact that the band is narrower than the interaction strength. However, coupling strength (weak or strong coupling) depends on d band width. According to values of Pd$_{X}$Ru$_{1-X}$ reported by Kusada et \textit{al.} \cite{Kusada2014} by using the Rietveld refinements of the XRD profiles, the lattice constant decreases linearly with increasing the molar ratio of the Ru. In this case, we can expect to get a stronger coupling for the Pd-rich NPs (with larger lattice constant) than the Ru-rich NPs (with smaller lattice constant). These results are consistent with the trend of the width trend from the Pd-NP (black curve in \textcolor{black}{Fig \ref{figure6} (b)}) to the Ru-NP (red curve in \textcolor{black}{Fig \ref{figure6} (d)}). 

There also exist some effects relative to the chemical nature of metals such as the number of electrons or the number of nearest neighbours \cite{Quaino2011}. We can underline that when the Ru ratio increase, totale  amount of electron change which can highly affect the charge transfer between the BM-NPs and the adsorbate molecule according to the model of Hammer and N??rskov \cite{Hammer2000}. From \textcolor{black}{Fig \ref{figure6} (c)}, the optimum catalytic activity of the Pd$_{0.5}$Ru$_{0.5}$-NP may result from the optimum balance between number of electron, the number of nearest neighbours and the spatial extension of the d-band (or orbitals). Indeed, even though the Pd$_{0.3}$Ru$_{0.7}$ and Pd$_{0.7}$Ru$_{0.3}$ have the same VBM that the Pd$_{0.5}$Ru$_{0.5}$, the CO oxidation of this latter is more efficient. Pd$_{0.5}$Ru$_{0.5}$-NP presents a large distribution of the electronic states (as shown by arrows in \textcolor{black}{Fig \ref{figure6} (c)}), which suggests that the electron back-donation (charge transfer) of the Pd can be tuned by the presence of the Ru. In fact, due to the d-band overlaps, electron can easily move from the Ru and Pd to the reactant.
\section*{Conclusions}
\indent \\
We analyzed chemical and electronic structures of Pd$_{X}$Ru$_{1-X}$ BM-NPs depending the Ru molar composition. We have employed the synchrotron-based hard X-ray photoelectron spectroscopy (HAXPES) to study the correlation between the chemisorption/catalytic performances and the core levels and valence band data. We found good agreement with the catalytic CO to CO$_{2}$ conversion behaviors and the physical-chemistry states of PdRu-NPs. C 1s core level spectra highlight the Pd$_{0.5}$Ru$_{0.5}$ NPs which provide the highest catalytic performance clearly reveals the presence of CO at the top site (preferential adsorption) whereas other samples exhibit two components associated with the bridge and hollow CO sites. We observe a higher adsorption on the Pd 3d$_{5/2}$ than Ru 3p$_{3/2}$ when the Ru content increase. Experimental results point out a transition from the alloy to the core-shell structures with increasing the Ru molar ratio. Optimum catalytic response of the Pd$_{0.5}$Ru$_{0.5}$-NP was related to the optimum d-band extension and most advantageous balance between the core-shell and alloy structures. An valuable consequence of our present study is that they provide the basics and the possibility to investigate future multi-metallic NPs Such as PdRuM (M : Ir, Au, Pt, etc.) and we hope that it will inspire such advances.


\subsection{Sample preparation}
\begin{figure*} 
\centering
\includegraphics[width=0.99\textwidth]{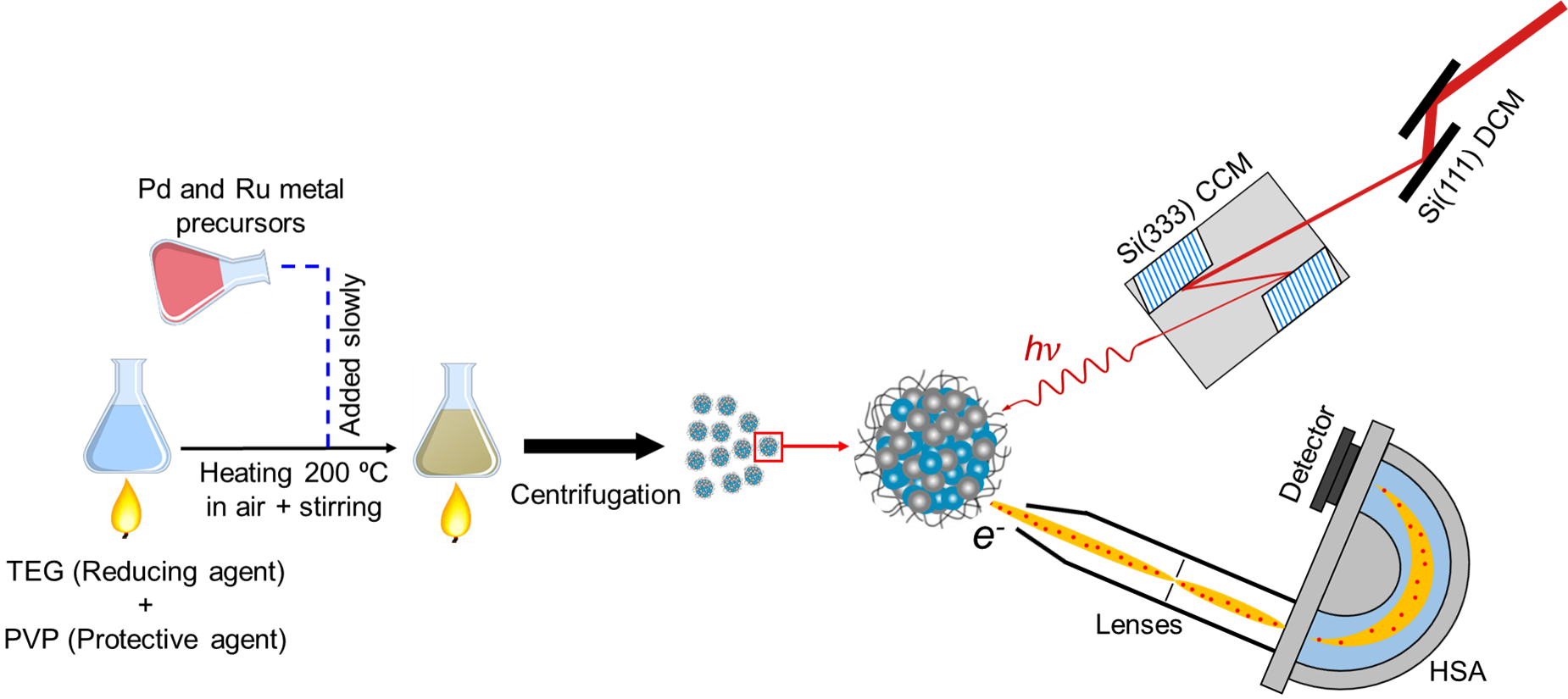}
\captionsetup{justification=raggedright} 
\captionof{figure}{Illustration showing the different steps from PdRu BM-NPs synthesis to HAXPES analyses (Not to scale). Left side: TEG and PVP stand for Triethylene glycol and Polyvinylpyrrolidone, respectively. Right side: DCM, CCM and HSA mean Double crystal monochromator, channel-cut crystal monochromator and Hemispherical Sector Analyzer, respectively.} 
\label{figure7}
\end{figure*}

Pd, Ru and Pd$_{X}$Ru$_{1-X}$ NPs ($X$=0, 0.1, 0.3, 0.5, 0.7, 0.9 and 1) samples were synthesized by using a chemical reduction method . More description of this technique can be found in ref \cite{Kusada2014}. Left part of  \textcolor{black}{Fig. \ref{figure7}} shows schematically a combination of Pd and Ru atoms. The poly(N-vinyl-2-pyrrolidone) (PVP) as the stabilizing agent was dissolved in triethylene glycol (TEG) used as the solvent and reducing agent. The solution was heated to 200$^{\circ}C$ in air with magnetic stirring. Simultaneously, metal precursors based on the K$_{2}$[PdCl$_{4}$] and RuCl$_{3}$.nH$_{2}$O were dissolved in deionized water and slowly added to the TEG solution. The solution was maintained at 200$^{\circ}C$  while adding the solution. After cooling down to room temperature, PdRu NPs were disassociated by centrifugation procedure. By tuning the Pd$^{2+}$/Ru$^{3+}$ molar ratio of the metal precursors, different nominal composition of binary NPs were obtained. The details of the size from Transmission electron microscopy observation and structures of PdRu NPs are listed in the \textcolor{black}{Table \ref{At_1}}.\\
\begin{table}
\centering
\caption{From the TEM images, sizes were estimated by averaging over at least 300 particles. The synchrotron XRD measurements were used to find the structures \cite{Kusada2014}.}
\label{At_1}
\begin{adjustbox}{max width=\textwidth}
\begin{tabular}{l c c ccc c cc}
\toprule
\midrule
Sample & & & & Mean diameters & & & & Patterns\\ 
\midrule
Pd & & & & 9.8 $\pm$2.6 & & & & fcc \\
Pd$_{0.9}$Ru$_{0.1}$ & & & &8.6  $\pm$1.4 & & & & fcc \\
Pd$_{0.7}$Ru$_{0.3}$ & & & &8.2 $\pm$1.6 & & & & fcc + hcp\\
Pd$_{0.5}$Ru$_{0.5}$ & & & &10.0 $\pm$1.2 & & & & fcc + hcp\\
Pd$_{0.3}$Ru$_{0.7}$ & & & &12.5 $\pm$2.2 & & & & fcc + hcp\\
Pd$_{0.1}$Ru$_{0.9}$ & & & &9.4 $\pm$ 1.7& & & & hcp\\
Ru & & & & 6.4 $\pm$ 1.7 & & & & hcp\\
\midrule 
\bottomrule
\end{tabular}
\end{adjustbox}
\end{table}

\subsection{Hard X-ray photoelectron spectroscopy analysis}
HAXPES measurements of Pd, Ru monometallic NPs as well as the PdRu bimetallic NPs were carried out at the National Institute for Materials Science (NIMS) contract beamline, BL15XU of the SPring-8 synchrotron facility in Japan  \cite{Ueda2013} as illustrated schematically in the right side of \textcolor{black}{Fig. \ref{figure7}}. A linearly polarized X-ray beam in the horizontal plane was used to investigate NPs. The excitation photon energy was established by using a Si 111 double crystal monochromator and the 333 reflection of a Si channel cut crystal monochromator.  Photoelectrons were analyzed by using a VG Scienta R4000 spectrometer. The total energy resolution (photon and spectrometer bandwidths) was estimated to be 0.231 eV (pass energy of 200 eV) for the excitation photon energy 5.95 keV. During the measurements, the analysis chamber pressure was maintained around 10$^{-8}$ mbar. HAXPES spectra were recorded at room temperature. Due to the insulated property of PVP, mono and bi-metallic NPs were mixed with carbon powder to cancelled the charge-up phenomenon. CasaXPS (CASA Software Ltd.) was used for data analysis \cite{CasaXPS2010}. The binding energy scale was calibrated by using the standard Au Fermi edge energy.

\bibliographystyle{natureprintstyle}

\begin{thebibliography}{1}

\bibitem{Mostafa2009}
S. Mostafa, J. R. Croy, H. Heinrich, B. R. Cuenya, Catalytic decomposition of alcohols over size-selected Pt nanoparticles supported on ZrO$_{2}$: A study of activity, selectivity, and stability, Appl. Catal. A 366, 353-362 (2009). 

\bibitem{Abad2008}
A. Abad, A. Corma and H. Garcia, Catalyst Parameters Determining Activity and Selectivity of Supported Gold Nanoparticles for the Aerobic Oxidation of Alcohols: The Molecular Reaction Mechanism, Chem. Eur. J. 14, 212-222 (2008).

\bibitem{Li2001}
Y. Li, M. A. El-Sayed, The Effect of Stabilizers on the Catalytic Activity and Stability of Pd Colloidal Nanoparticles in the Suzuki Reactions in Aqueous Solution, J. Phys. Chem. B 105, 8938-8943 (2001).

\bibitem{Gasteiger2009}
H. A. Gasteiger, N. M. Markovi, Just a Dream-or Future Reality?, Science 324, 48-49 (2009).

\bibitem{Weskamp1998}
T. Weskamp, W. C. Schattenmann, M. Spiegler, W. A. Herrmann, A novel class of ruthenium catalysts for olefin metathesis, Angew. Chem., Int. Ed. 37, 2490-2493 (1998).

\bibitem{Yamaguchi2003}
K. Yamaguchi, N. Mizuno, Efficient heterogeneous aerobic oxidation of amines by a supported ruthenium catalyst, Angew. Chem. Int. Ed. 42, 1480-1483 (2003).

\bibitem{Jacobsen2000}
C. J. H. Jacobsen et al., Structure sensitivity of supported ruthenium catalysts for ammonia synthesis, J. Mol. Catal. A-Chem 163, 19-26 (2000).

\bibitem{Perkas2011}
N. Perkas et al., Supported Ru catalysts prepared by two sonication-assisted methods for preferential oxidation of CO in H$_{2}$, Phys. Chem. Chem. Phys. 13, 15690-15698 (2011).

\bibitem{Bowker2008}
M. Bowker, Automotive catalysis studied by surface science, Chem. Soc. Rev. 37, 2204-2211 (2008).

\bibitem{Singh2013}
A. K. Singh and Q. Xu, Synergistic catalysis over bimetallic alloy nanoparticles, ChemCatChem 5, 652-676 (2013).

\bibitem{Kim2014}
D. Kim, J. Resasco, Y. Yu, A. M. Asiri, P. Yang, Synergistic geometric and electronic effects for electrochemical reduction of carbon dioxide using gold-copper bimetallic nanoparticles, Nat. Commun. 5, 4948 (2014).

\bibitem{Jiang2011}
H. L. Jiang, T. Akita, T. Ishida, M. Haruta, Q. Xu, Synergistic catalysis of Au@Ag core-shell nanoparticles stabilized on metal-organic framework, J. Am. Chem. Soc. 133, 1304-1306 (2011).

\bibitem{Mott2007}
D. Mott et al., Synergistic activity of gold-platinum alloy nanoparticle catalysts, Catal. Today 122, 378-385 (2007).

\bibitem{Toshima1998}
N. Toshima, T. Yonezawa, Bimetallic nanoparticles-novel materials for chemical and physical applications, New J. Chem. 1179-1201 (1998).

\bibitem{Kusada2010}
K. Kusada, M. Yamauchi, H. Kobayashi, H. Kitagawa and Y. Kubota, Hydrogen-storage properties of solid-solution alloys of immiscible neighboring elements with Pd, J. Am. Chem. Soc. 132, 15896-15898 (2010).

\bibitem{Kusada2014}
K. Kusada, et al., Solid Solution Alloy Nanoparticles of Immiscible Pd and Ru Elements Neighboring on Rh: Changeover of the Thermodynamic Behavior for Hydrogen Storage and Enhanced CO-Oxidizing Ability, J. Am. Chem. Soc. 136, 1864-1871 (2014).

\bibitem{Wu2014}
D. Wu, M. Cao, M. Shen and R. Cao, Sub-5 nm Pd-Ru nanoparticle alloys as efficient catalysts for formic acid electrooxidation, ChemCatChem 6, 1731-1736 (2014).

\bibitem{Wu2016}
D. Wu, K. Kusada, H. Kitagawa, Recent progress in the structure control of Pd-Ru bimetallic nanomaterials, Sci. Technol. Adv. Mater. 17, 583-596 (2016).

\bibitem{Sasaki2012}
K. Sasaki et al., Highly stable Pt monolayer on PdAu nanoparticle electrocatalysts for the oxygen reduction reaction, Nat. Commun. 3, 1115 (2012).

\bibitem{Liu2004}
Z. Liu, X. Y. Ling, X. Su and J. Y. Lee, Carbon-supported Pt and PtRu nanoparticles as catalysts for a direct methanol fuel cell, J. Phys. Chem. B 108, 8234-8240 (2004).

\bibitem{Alayoglu2008}
S. Alayoglu, A. U. Nilekar, M. Mavrikakis and B. Eichhorn, Ru-Pt core-shell nanoparticles for preferential oxidation of carbon monoxide in hydrogen, Nat. Mater. 7, 333-338 (2008).

\bibitem{Adekoya2015}
J. A. Adekoya, E. O. Dare, M. A. Mesubi and N. Revaprasadu, Facile route to the synthesis and characterization of novel core-shell and Ag/Ru allied nanoparticles, Phys. E 71, 70-78 (2015).

\bibitem{Esfahani2010}
H. A. Esfahani, L. Wang, Y. Nemoto and Y. Yamauchi, Synthesis of bimetallic Au@Pt nanoparticles with Au core and nanostructured Pt shell toward highly active electrocatalysts, Chem. Mater. 22, 6310-6318 (2010).

\bibitem{Song2019}
C. Song et al., Correlation between the electronic/local structure and CO-oxidation activity of Pd$ _{x} $Ru$ _{1-x} $ alloy nanoparticles, Nanoscale Adv. 1, 546-553 (2019).

\bibitem{Guidez2012}
E. B. Guidez, V. M??kinen, H. H??kkinen and C. M. Aikens, Effects of Silver Doping on the Geometric and Electronic Structure and Optical Absorption Spectra of the Au$ _{25-n} $Ag$ _{n} $(SH)$ _{18} $-(n = 1, 2, 4, 6, 8, 10, 12) Bimetallic Nanoclusters, J. Phys. Chem. C 116, 20617-20624 (2012).

\bibitem{McFarland2012}
E. McFarland, Unconventional chemistry for unconventional natural gas, Science 338, 340-342 (2012).

\bibitem{Lee2012}
H. C. Lee, Y. Potapova and D. Lee, A core-shell structured, metal-ceramic composite-supported Ru catalyst for methane steam reforming, J. Power Sources 216, 256-260 (2012).

\bibitem{Xiao2009}
L. Xiao, L. Zhuang, Y. Liu, J. Lu and H. D. Abruna, Activating Pd by morphology tailoring for oxygen reduction, J. Am. Chem. Soc. 131, 602-608 (2009).

\bibitem{Newton2007}
M. A. Newton, C. Belver-Coldeira, A. Martinez-Arias, M. Fernandez-Garcia, Dynamic in situ observation of rapid size and shape change of supported Pd nanoparticles during CO/NO cycling, Nat. Mater. 6, 528-532 (2007).

\bibitem{Schultz2006}
N. E. Schultz, B. E. Gherman, C. J. Cramer, D. G. Truhlar, Pd$ _{n} $CO (n = 1,2): Accurate Ab Initio Bond Energies, Geometries, and Dipole Moments and the Applicability of Density Functional Theory for Fuel Cell Modeling, J. Phys. Chem. B 110, 24030-24046 (2006).

\bibitem{Gueye2019}
I. Gueye et al., Journal of Catalysis, Investigation of selective chemisorption of fcc and hcp Ru nanoparticles using X-ray photoelectron spectroscopy analysis, doi:10.1016/j.jcat.2019.10.004

\bibitem{Chorkendorff2003}
I. Chorkendorff, J. W. Niemantsverdriet, Concepts of Modern Catalysis and Kinetics; Wiley-VCH: New York, 2003.

\bibitem{Allian2011}
A. D. Allian et al., Chemisorption of CO and Mechanism of CO Oxidation on Supported
Platinum Nanoclusters, J. Am. Chem. Soc. 133, 12, 4498-4517 (2011).

\bibitem{LaRue2015}
J. L. LaRue et al., THz-pulse-induced selective catalytic CO oxidation on Ru, Phys. Rev. Lett. 115, 036103 (2015).

\bibitem{Kaichev2003}
V. V. Kaichev et al., High-pressure studies of CO adsorption on Pd (111) by X-ray photoelectron spectroscopy and sum-frequency generation, J. Phys. Chem. B 107, 3522-3527 (2003).

\bibitem{Naitabdi2018}
A. Naitabdi et al., CO oxidation activity of Pt, Zn and ZnPt nanocatalysts: a comparative study by in situ near-ambient pressure X-ray photoelectron spectroscopy, Nanoscale 10, 6566-6580 (2018).

\bibitem{Schiffer1997}
A. Schiffer, P. Jacob, D. Menzel, The (2CO+O)(2x2) Ru (001) layer: preparation, characterization, and analysis of interaction effects in vibrational spectra, Surf. Sci. 389, 116-130 (1997).

\bibitem{Narloch1995}
B. Narloch, G. Held,  D. Menzel, A LEED-IV determination of the Ru (001)-p (2x2)(O+CO) structure: A coadsorbate-induced molecular tilt, Surf. Sci. 340, 159-171 (1995).

\bibitem{Kostov1992}
K. L. Kostov, H. Rauscher, D. Menzel, Adsorption of CO on oxygen-covered Ru (001), Surf. Sci. 278, 62-86 (1992).

\bibitem{Westerstrom2011}
R. Westerstr{\"o}m et al, Oxidation and reduction of Pd (100) and aerosol-deposited Pd nanoparticles, Phys. Rev. B 83, 115440 (2011).

\bibitem{Toyoshima2012}
R. Toyoshima et al, Active surface oxygen for catalytic CO oxidation on Pd (100) proceeding under near ambient pressure conditions, J. Phys. Chem. Lett. 3, 3182-3187 (2012).

\bibitem{Morgan2015}
D. J. Morgan, Resolving ruthenium: XPS studies of common ruthenium materials, Surf. Interface Anal. 47, 1072-1079 (2015).

\bibitem{Hota2019}
M. K. Hota et al, Integration of Electrochemical Microsupercapacitors with Thin Film Electronics for On-Chip Energy Storage, Adv. Mater. 31, 1807450 (2019).

\bibitem{Kruefu2016}
V. Kruefu et al, Enhancement of p-type gas-sensing performances of NiO nanoparticles prepared by precipitation with RuO$ _{2} $ impregnation, Sensor. Actuat. B-Chem. 236, 466-473 (2016).

\bibitem{Lizzit2009}
S. Lizzit et al, O-and H-induced surface core level shifts on Ru (0001): prevalence of the additivity rule, J. Phys.: Condens. Matter 21, 134009 (2009).

\bibitem{Andersson2009}
K. J. Andersson, F. Calle-Vallejo, J. Rossmeisl, I. Chorkendorff, Adsorption-driven surface segregation of the less reactive alloy component, J. Am. Chem. Soc. 131, 6, 2404-2407 (2009).

\bibitem{Knudsen2007}
J. Knudsen et al, A Cu/Pt Near-Surface Alloy for Water-Gas Shift Catalysis,  J. Am. Chem. Soc. 129, 6485-6490 (2007).

\bibitem{Nierhoff2015}
A. Nierhoff et al, Adsorbate induced surface alloy formation investigated by near ambient pressure X-ray photoelectron spectroscopy, Cataly. Today 244, 130-135 (2015).

\bibitem{Alayoglu2009}
S. Alayoglu et al, Structural and architectural evaluation of bimetallic nanoparticles: a case study of Pt-Ru core-shell and alloy nanoparticles, ACS Nano 3, 3127-3137 (2009).

\bibitem{Miyakawa2014}
M. Miyakawa et al, Continuous syntheses of Pd@Pt and Cu@Ag core-shell nanoparticles using microwave-assisted core particle formation coupled with galvanic metal displacement,  Nanoscale 6, 8720-8725 (2014).

\bibitem{Peljo2016}
P. Peljo, J. A. Manzanares, H. H. Girault, Contact potentials, fermi level equilibration, and surface charging, Langmuir 32, 5765-5775 (2016).

\bibitem{Kumsa2016}
D. W. Kumsa et al, Electron transfer processes occurring on platinum neural stimulating electrodes: a tutorial on the i(V$ _{e} $) profile, J. Neural Eng. 13, 052001 (2016).

\bibitem{Gueye2016}
I. Gueye et al, Chemistry of surface nanostructures in lead precursor-rich PbZr$ _{0. 52} $Ti$ _{0. 48} $O$ _{3} $ sol-gel films, Appl. Surf. Sci. 363, 21-28 (2016).

\bibitem{Newns1969Grimley1971}
D. M. Newns, Self-consistent model of hydrogen chemisorption, Phys. Rev. 178, 1123-1135 (1969); T. B. Grimley, Electronic structure of adsorbed atoms and molecules, J. Vac. Sci. Techno. 8, 31-38 (1971).

\bibitem{Quaino2011}
P. Quaino, E. Santos, G. Soldano and W. Schmickler, Recent Progress in Hydrogen Electrocatalysis, Adv. Phys. Chem. 2011, 851640 (2011).

\bibitem{Hammer2000}
B. Hammer,  J. K. N$ \phi $rskov, Theoretical surface science and catalysis-calculations and concepts, Adv. Catal. 45, 71-129 (2000).

\bibitem{Ueda2013}
S. Ueda et al., Present Status of the NIMS Contract Beamline BL15XU at SPring-8, AIP Conf. Proc. 1234, 403 (2010).

\bibitem{CasaXPS2010}
J. Walton, P. Wincott, N. Fairley, A. Carrick, Peak Fitting with
CasaXPS: A Casa Pocket Book, Acolyte Science, Cheshire, 2010. 
\end{thebibliography}

\indent \\
\textbf{Author acknowledgements}\\
This work was partly supported by ACCEL, Japan Science and Technology Agency (JST) under Grant No. JPMJAC1501. The measurements were performed at SPring-8 with the approval of the NIMS synchrotron X-ray Station at SPring-8 under proposal Nos. 2013A4908, as part of the project NIMS Nanotechnology Platform (Project No. 12024046, Proposal No. A-13-NM-0009), 2018A4506, and 2018B4504, and with the approval of the
JASRI under proposal No. 2018A1351. This work was also partly supported by the Ministry of Education, Culture, Sports, Science and Technology of Japan (OS:18K04868).\\
\textbf{Author contributions}\\
O. Sakata, I. Gueye and H. Kitagawa managed the project. 
K. Kusada and H. Kitagawa synthesized the samples. 
O. Sakata, A. Yang, L. S. R. Kumara performed the HAXPES experiments.   
I. G. analysed all the experimental data. I. Gueye and O. Sakata. wrote the manuscript with contribution from S. Hiroi, L. S. R. Kumara, O. Seo and J. Kim.\\
\textbf{Competing Interests}\\
The authors declare that they have no competing interests.\\
\textbf{Data and materials availability}\\
The datasets that support the findings of the current study are available from the corresponding author on reasonable request.\\

\end{document}